\documentclass[twocolumn,showpacs,aps,floatfix,superscriptaddress]{revtex4}
\usepackage{amsmath,amssymb,graphicx,eucal,bm}
\begin{document}
\title{Kinetics of First Passage in a Cone}
\author{E.~Ben-Naim}
\affiliation{Theoretical Division and Center for Nonlinear
Studies, Los Alamos National Laboratory, Los Alamos, New Mexico
87545, USA}
\author{P.~L.~Krapivsky}
\affiliation{Department of Physics,
Boston University, Boston, Massachusetts 02215, USA}
\begin{abstract}
We study statistics of first passage inside a cone in arbitrary
spatial dimension.  The probability that a diffusing particle avoids
the cone boundary decays algebraically with time.  The decay exponent
depends on two variables: the opening angle of the cone and the
spatial dimension.  In four dimensions, we find an explicit expression
for the exponent, and in general, we obtain it as a root of a
transcendental equation involving associated Legendre functions.  At
large dimensions, the decay exponent depends on a single scaling
variable, while roots of the parabolic cylinder function specify the
scaling function.  Consequently, the exponent is of order one only if
the cone surface is very close to a plane.  We also perform asymptotic
analysis for extremely thin and extremely wide cones.
\end{abstract}
\maketitle

\section{Introduction}

Random walks are widely used to model natural processes in physics,
chemistry, and biology \cite{wf,ghw,hcb,rg}. In particular, 
first-passage and persistence statistics \cite{sr,snm} of multiple
random walks underlie reaction-diffusion processes \cite{kbr}, spin
systems \cite{dhp,msbc,dhz}, and polymer dynamics \cite{bdve,is}.

First-passage processes involving multiple random walks are equivalent
to diffusion in a restricted region of space.  For example, the
probability that $N$ ordinary random walks do not meet is equivalent
to the probability that a ``compound'' walk in $N$ dimensions remains
confined to the region \hbox{$x_1<x_2<\ldots<x_N$}. This probability
decays as \hbox{$t^{-N(N-1)/4}$} in the long-time limit
\cite{mef,hf,djg}.

When there are only two or three particles, the compound walk is, in
many cases, confined to a wedge, formed by two intersecting
planes. Moreover, the well-known properties of diffusion inside an
absorbing wedge \cite{sr} explain the long-time kinetics
\cite{mef,fg,dba,kr,bjmkr}.  In general, however, the absorbing
boundary is defined by multiple intersecting planes in a
high-dimensional space. Apart from a few special cases, diffusion
subject to such complicated boundaries conditions remains an open
problem \cite{hn,bg,kr,bjmkr,ck}.

Our goal is to use cones in high dimensions to approximate the
absorbing boundaries that underlie such first-passage processes.  In
this study, we obtain analytic results for the survival probability of
a diffusing particle inside an absorbing cone in arbitrary dimension.
In a follow-up study \cite{bk}, we demonstrate that cones provide
useful approximations to first-passage characteristics of multiple
random walks \cite{yal}.

\begin{figure}[t]
\includegraphics[width=0.25\textwidth]{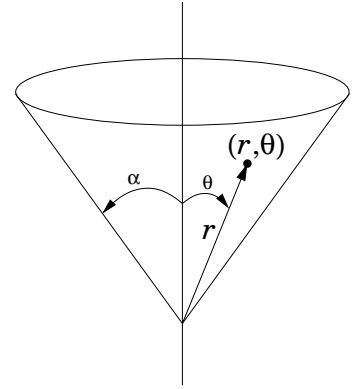}
\caption{Illustration of a cone with opening angle $\alpha$. The
initial location of the particle is parametrized by the radial
distance $r$ and the polar angle $\theta$.}
\label{fig-cone}
\end{figure}

We consider a single particle that diffuses inside an unbounded cone
with opening angle $\alpha$ in spatial dimension $d$ (Figure
\ref{fig-cone}). The central quantity in our study is the probability
$S(t)$ that the particle does not reach the cone boundary up to time
$t$. Regardless of the starting position, this survival probability
decays algebraically, $S\sim t^{-\beta}$, in the long-time limit.

First, we find the exponent $\beta$ analytically by solving the
Laplace equation inside the cone.  In dimensions two and four, this
exponent is an explicit function of the opening angle $\alpha$, and in
particular, $\beta=(\pi-\alpha)/2\alpha$ when $d=4$. In general
dimension, we find $\beta$ as a root of a transcendental equation
involving the associated Legendre functions.

Second, we derive scaling properties of the exponent.  Interestingly,
the exponent $\beta$ becomes a function of a single scaling variable
in the large-$d$ limit.  We obtain the scaling function as a root of
the transcendental equation
\begin{equation}
D_{2\beta}(y)=0\quad {\rm with}\quad y=(\cos\alpha)\sqrt{d}
\end{equation}
involving the parabolic cylinder function $D_\nu$.  The exponent
$\beta$ is of order one only in a small region around
$\alpha=\pi/2$. The width of this region shrinks as $d^{-1/2}$ in the
infinite dimension limit. The exponent diverges algebraically,
$\beta(y)\simeq y^2/8$ as $y\to \infty$, and it is exponentially
small, $\beta(y)\simeq \sqrt{y^2/8\pi}\exp(-y^2/2)$ when $y\to
-\infty$. Thus, in the large-$d$ limit, the exponent $\beta$ is huge
if the opening angle is acute, and conversely, it is tiny if the
opening angle is obtuse. Strikingly, if we fix the opening angle
$\alpha$ and take the limit $d\to\infty$, there are three distinct
possibilities,
\begin{equation}
\label{three}
\lim_{d\to\infty}\beta_d(\alpha)=
\begin{cases}
\infty&\alpha<\pi/2,\\
1/2&\alpha=\pi/2,\\
0&\alpha>\pi/2.\\
\end{cases}
\end{equation}
Of course, a cone with opening angle $\alpha=\pi/2$ is simply a plane,
and hence, $\beta(\alpha=\pi/2)=1/2$ for all $d$.

Third, we study the limiting cases of very thin and very wide cones.
The exponent diverges algebraically, \hbox{$\beta\sim \alpha^{-1}$},
when the cone is extremely thin. When the cone is extremely wide, the
exponent is exponentially small, \hbox{$\beta\sim
(\pi-\alpha)^{d-3}$}.

The rest of this paper is organized as follows. In Section II, we
write the diffusion equation that governs the survival probability,
and show that finding the leading asymptotic behavior of the survival
probability requires a solution to the Laplace equation
\cite{rdd,dz,bs,bd,ntv}.  We present the solutions to this Laplace
equation in two and four dimensions in Section III, and for an
arbitrary dimension in Section IV. The bulk of the paper deals with
asymptotic analysis for very large dimensions.  In particular, we
derive scaling properties of the exponent and obtain the limiting
behaviors of the scaling function (Section V).  Asymptotic results for
extremely thin and extremely wide cones are detailed in Sections VI
and VII, respectively. We also obtain the first-passage time (Section
VIII) and conclude with a discussion in Section IX.

\section{The Diffusion Equation}

Consider a particle undergoing Brownian motion \cite{bdup,mp} inside
an unbounded cone in spatial dimension $d$.  The opening angle
$\alpha$, that is, the angle between the cone axis and its surface,
fully specifies the cone (Figure \ref{fig-cone}). The range of opening
angles is $0\leq \alpha\leq \pi$, and for $\alpha=\pi/2$, the cone
surface is planar.  Moreover, the exterior of the cone is itself a
cone with opening angle $\pi-\alpha$.  In two dimensions, the cone is
a wedge, and in three dimensions, the
cone is an ordinary circular cone.

At time $t=0$, the particle is released from a certain location inside
the cone. Our goal is to determine the probability that the particle
does not reach the cone surface up to time $t$.  By symmetry, this
survival probability, $S\equiv S(r,\theta,t)$, depends on the initial
distance to the apex $r$, and the initial angle with the cone axis
$\theta$. Using a spherical coordinate system where the origin is
located at the cone apex and the $z$-axis is along the cone axis, the
pair of parameters $(r,\theta)$ are simply the radial and the polar
angle coordinates of the initial location (Figure \ref{fig-cone}).

The survival probability fully quantifies the first-passage process.
For example, the probability that the particle first reaches the cone
surface during the time interval \hbox{$(t,t+dt)$} equals
$[-dS(r,\theta,t)/dt]\times dt$.  In general, the survival probability
satisfies the diffusion equation \cite{ghw}
\begin{equation}
\label{S-eq}
\frac{\partial S(r,\theta,t)}{\partial t}=D\nabla^2 S(r,\theta,t),
\end{equation}
where $D$ is the diffusion constant.  The initial condition is
$S(r,\theta,t=0)=1$, and the boundary condition is $S(r,\alpha,
t)=S(0,\theta,t)=0$.

We are primarily interested in the large-time kinetics. Based on the
behavior in two and three dimensions \cite{sr}, we expect that the
survival probability decays algebraically,
\begin{equation}
\label{decay}
S(r,\theta,t) \simeq \,\Phi(r,\theta)\, t^{-\beta},
\end{equation}
as $t\to\infty$. The exponent $\beta\equiv \beta_d(\alpha)$ depends on
the spatial dimension $d$ and the opening angle $\alpha$. The
dependence on the initial location enters only through the amplitude
$\Phi$.

We now substitute the leading asymptotic behavior \eqref{decay} into
the diffusion equation \eqref{S-eq}. Since the time derivative becomes
negligible in the long-time limit, the amplitude $\Phi$ satisfies
Laplace's equation, 
\begin{equation}
\label{Phi-eq}
\nabla^2 \Phi(r,\theta)=0,
\end{equation}
subject to the boundary conditions
\hbox{$\Phi(r,\alpha)=\Phi(0,\theta)=0$}. The survival probability is
finite and positive everywhere inside the cone, and consequently, the
amplitude must be finite and positive, $0<\Phi(r,\theta)<\infty$ for
all $\theta<\alpha$. In addition,
\hbox{$d\Phi/d\theta|_{\theta=0}=0$}, to avoid a cusp along the cone
axis.

We use dimensional analysis \cite{krb} to solve
Eq.~\eqref{Phi-eq}. The survival probability is dimensionless, and
hence, \eqref{decay} implies $[\Phi]=T^\beta$ where $T$ denotes
dimension of time.  The amplitude $\Phi$ depends on three variables:
the radial coordinate $r$ with dimension of length, $[r]=L$, the
diffusion coefficient $D$ with $[D]=L^2/T$, and the dimensionless
angle $\theta$.  As the quantity $(r^2/D)^\beta$ is the only
combination of the variables $r$ and $D$ with dimension $T^\beta$, we
seek a solution in the form
\begin{equation}
\label{Phi-sep}
\Phi(r,\theta)=\left(\frac{r^2}{D}\right)^{\beta}\psi(\theta).
\end{equation}
The angular function $\psi(\theta)$ must be finite and positive
everywhere inside the cone, $0<\psi(\theta)<\infty$ for
$\theta<\alpha$, and it vanishes on the cone surface,
$\psi(\alpha)=0$.

We now write the Laplacian operator in \eqref{Phi-eq} explicitly by
using
\begin{equation*}
\nabla^2 \equiv \frac{\partial^2}{\partial r^2}\!+
\!\frac{d\!-\!1}{r}\frac{\partial}{\partial r}\!+\!
\frac{1}{r^2(\sin\theta)^{d\!-\!2}}\frac{\partial}{\partial
\theta}(\sin\theta)^{d\!-\!2} \frac{\partial}{\partial\theta}.
\end{equation*}
Next, we substitute \eqref{Phi-sep} into the Laplace equation and find
that the angular function $\psi\equiv \psi(\theta)$ satisfies the
eigenvalue equation \cite{rdd,dz,bs,bd,ntv}
\begin{equation}
\label{psi-theta-eq}
\frac{1}{(\sin\theta)^{d-2}}\frac{d}{d \theta}\left[(\sin\theta)^{d-2}
\frac{d\psi}{d  \theta}\right]+2\beta(2\beta+d-2)\psi=0. 
\end{equation}
The boundary conditions are $\psi'(0)=0$ and $\psi(\alpha)=0$.

From equations \eqref{decay} and \eqref{Phi-sep}, the leading
asymptotic behavior is
\begin{equation}
\label{decay1}
S(r,\theta,t) \simeq
\psi(\theta)\,\left(\frac{D\,t}{r^2}\right)^{-\beta},
\end{equation}
as $t\to\infty$.  In particular, the survival probability grows
algebraically with distance, $S\simeq r^{2\beta}$.  We note that the
problem of finding the leading asymptotic behavior of the survival
probability reduces to an electrostatic problem \cite{cj,jdj} as the
amplitude satisfies Laplace's equation \eqref{Phi-eq}.

Our goal is to find the exponent $\beta$ and the angular function
$\psi(\theta)$. We expect that the exponent is a monotonically
decreasing function of the opening angle $\alpha$. Also, in all
dimensions, $\beta(\alpha=\pi/2)=1/2$ because a cone with opening
angle $\alpha=\pi/2$ is a half-space and consequently, the
first-passage probability is identical to that of a particle in the
vicinity of a trap in one dimension \cite{sr}.

\section{Dimensions Two and Four}

We first discuss two special cases where explicit solutions are
possible.  In dimension two, the second order differential equation
\eqref{psi-theta-eq} reads
\hbox{$\psi_{\theta\theta}+(2\beta)^2\,\psi=0$}, and the two
independent solutions are simply $\cos(2\beta\theta)$ and
$\sin(2\beta\theta)$.  The boundary condition $\psi'(0)=0$ excludes
the latter and therefore,
\begin{equation}
\label{psi2}
\psi_2(\theta)=\cos(2\beta\theta).
\end{equation}
Henceforth, the subscript indicates the dimension. Since the linear
equation \eqref{psi-theta-eq} specifies the angular function only up
to an overall constant, we set the prefactor to one throughout this
paper.  The boundary condition \hbox{$\psi(\alpha)=0$} and the
requirement that the angular function must be positive everywhere
inside the cone together imply $2\beta\alpha=\pi/2$. Therefore, the
exponent is \cite{sr}
\begin{equation}
\label{beta2}
\beta_2(\alpha)=\frac{\pi}{4\alpha}.
\end{equation}
As expected, the exponent is a monotonically decreasing function of
$\alpha$, and $\beta_2(\pi/2)=1/2$. The exponent is minimal, yet
finite, $\beta_2(\pi)=1/4$, for an absorbing needle \cite{cr}. Thus, in
dimension two, a diffusing particle reaches a needle with certainty.

In dimension four, the transformation
\hbox{$\psi_4(\theta)=(\sin\theta)^{-1}u(\theta)$} reduces the
eigenvalue equation \eqref{psi-theta-eq} to
$u_{\theta\theta}+(2\beta+1)^2u=0$.  Now, the two independent
solutions are $\sin[(2\beta+1)\theta]$ and $\cos[(2\beta+1)\theta]$,
but the latter is forbidden because the function $\psi$ must be
finite.  Therefore,
\begin{equation}
\label{psi4}
\psi_4(\theta)= \frac{\sin\left[(2\beta+1)\theta\right]}{\sin\theta}.
\end{equation}
The boundary condition $\psi(\alpha)=0$ and the requirement
$\psi(\theta)>0$ for $\theta<\alpha$ together give 
$(2\beta+1)\alpha=\pi$. Hence, the exponent is an explicit function of
the opening angle in dimension four as well,
\begin{equation}
\label{beta4}
\beta_4(\alpha)=\frac{\pi-\alpha}{2\alpha}.
\end{equation}
The exponent vanishes, $\beta_4(\alpha)\to 0$ as $\alpha\to \pi$, so
it is no longer guaranteed that a diffusing particle reaches a needle.

\section{General Dimension}

In general dimension, we transform the eigenvalue equation
\eqref{psi-theta-eq} using the variable \hbox{$\mu=\cos\theta$}. In terms of
this variable, the function $\psi\equiv \psi_d(\mu)$ satisfies
\begin{equation}
\label{psi-mu-eq}
(1-\mu^2)\frac{d^2\psi}{d\mu^2}-(d-1)\,\mu\,\frac{d\psi}{d\mu}
+2\beta(2\beta+d-2)\psi=0.
\end{equation}
The boundary condition is $\psi(\cos\alpha)=0$. We now use a second
transformation,
\begin{equation*}
\psi(\mu)=\left(1-\mu^2\right)^{-\delta/2}\!\Psi(\mu) \quad {\rm with}\quad
\delta=\frac{d-3}{2}.
\end{equation*}
Substituting this form into \eqref{psi-mu-eq} shows that the auxiliary
function $\Psi\equiv \Psi_d(\mu)$ satisfies the associated Legendre
equation \cite{as,NIST}
\begin{eqnarray}
\label{leg-eq}
(1-\mu^2)\frac{d^2\Psi}{d\mu^2}&-&2\,\mu\,\frac{d\Psi}{d\mu}\\
\nonumber
&+&\left[(2\beta+\delta)(2\beta+\delta+1)-\frac{\delta^2}{1-\mu^2}\right]
\Psi=0.
\end{eqnarray}
The two independent solutions of this equation are
$P_{2\beta+\delta}^\delta(\mu)$ and $Q_{2\beta+\delta}^\delta(\mu)$,
the associated Legendre functions of degree $2\beta+\delta$ and order
$\delta$.  In even dimensions, the first solution is not physical
because it implies a divergence of $\psi$ as $\theta\to 0$
\cite{NIST}.  In odd dimensions, the second solution is excluded for
the same reason.  Therefore \cite{note},
\begin{equation}
\label{psid}
\psi_d(\theta)=
\begin{cases}
(\sin\theta)^{-\delta}P_{2\beta+\delta}^\delta(\cos\theta)& d\ {\rm odd},\\
(\sin\theta)^{-\delta}Q_{2\beta+\delta}^\delta(\cos\theta)& d\ {\rm even}.
\end{cases}
\end{equation}
The boundary condition $\psi(\alpha)=0$ relates the exponent $\beta$
and the opening angle $\alpha$,
\begin{equation}
\label{betad}
\begin{split}
P_{2\beta+\delta}^\delta(\cos\alpha) &= 0\qquad d\ {\rm odd},\\
Q_{2\beta+\delta}^\delta(\cos\alpha) &= 0\qquad d\ {\rm even}.
\end{split}
\end{equation}
In general dimension, the exponent $\beta$ is the smallest root of the
transcendental equation \eqref{betad} involving the associated
Legendre functions. We must always choose the smallest root because
$\psi(\theta)>0$ for all $\theta<\alpha$.

\begin{figure}[t]
\includegraphics[width=0.45\textwidth]{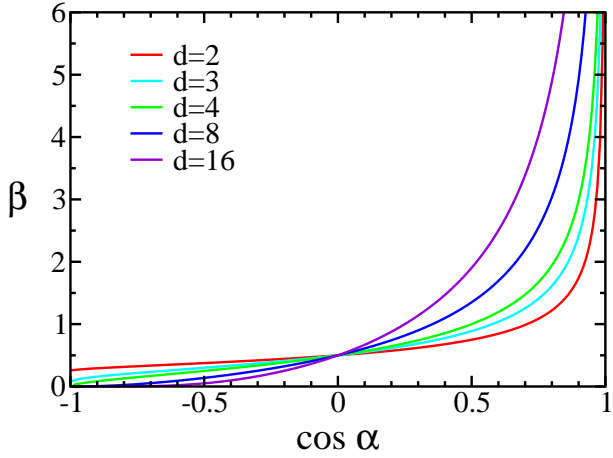}
\caption{The survival exponent $\beta$ versus $\cos\alpha$ at
different dimensions. This exponent is given explicitly by
\eqref{beta2} and \eqref{beta4} in dimensions $d=2$ and $d=4$,
respectively. In general, the exponent is a root of the transcendental
equation \eqref{betad}.}
\label{fig-beta}
\end{figure}

We can verify that the exponent $\beta$ decreases monotonically with
the opening angle $\alpha$, and that $\beta(\pi/2)=1/2$ (Figure
\ref{fig-beta}).  In all dimensions, the exponent diverges for
extremely thin cones, $\alpha\to 0$, and when $d\geq 3$, the exponent
vanishes when $\alpha\to \pi$. 

\begin{figure}[t]
\includegraphics[width=0.4\textwidth]{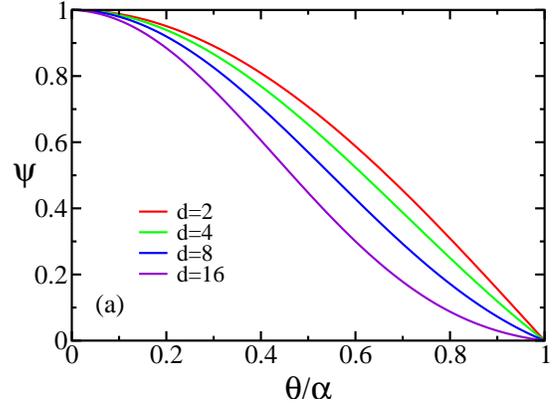}\vspace{.3in}
\includegraphics[width=0.4\textwidth]{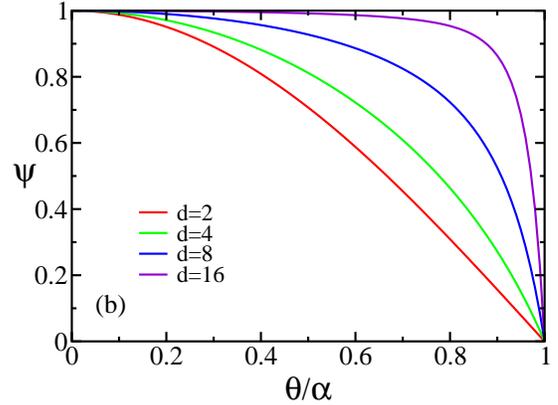}
\caption{The function $\psi(\theta)$ versus the normalized polar angle
$\theta/\alpha$ at different dimensions. The function $\psi(\theta)$
is normalized such that $\psi(0)=1$.  Figure 3a shows the behavior for
an acute opening angle, $\alpha=\pi/4$, and Figure 3b shows the
behavior for an obtuse opening angle, $\alpha=3\pi/4$.}
\label{fig-psi}
\end{figure}

Let us fix the opening angle and increase the dimension. There are two
possibilities: (i) if $\alpha<\pi/2$, the first-passage process speeds
up with increasing dimension because $\beta$ increases (Figure
\ref{fig-beta}) and $\psi(\theta)$ declines (Figure \ref{fig-psi}a);
(ii) if $\alpha>\pi/2$, the first-passage process slows down because
$\beta$ shrinks (Figure \ref{fig-beta}) and $\psi(\theta)$ grows
(Figure \ref{fig-psi}b). Ultimately, in the infinite-dimension limit,
first passage becomes instantaneous for acute angles but infinitely
long for obtuse angles.

In dimension three, $\delta=0$, and the solution \eqref{psid} reduces to the
Legendre function of index $2\beta$, namely, \hbox{$\psi_3(\theta)=
P_{2\beta}(\cos\theta)$} \cite{as,NIST}.  Also, the exponent
$\beta$ is the smallest root of the transcendental equation \cite{sr}
\begin{equation}
\label{beta3}
P_{2\beta}(\cos\alpha)=0.
\end{equation}

For half-integer values of the exponent, $\beta=n/2$, the angular
function $\psi_d(\theta)$ is a polynomial of degree $n$ in
$\cos\theta$. This follows directly from equation
\eqref{psi-mu-eq}. For example,
\begin{equation}
\label{psi-special}
\psi_d(\theta)=
\begin{cases}
\cos\theta&\beta=1/2,\\
d\cos^2\theta-1&\beta=1,\\
(d+2)\cos^3\theta-3\cos\theta&\beta=3/2.
\end{cases}
\end{equation}
In three dimensions, the polynomials coincide with the Legendre
polynomials: $\psi_3(\theta)=P_n(\cos\theta)$ when
\hbox{$\beta=n/2$}.  Using $\psi(\alpha)=0$, we find the opening
angles for which the exponent is a half-integer:
\begin{equation}
\label{beta-special}
{\beta}_d(\alpha)=
\begin{cases}
1/2\quad&\cos\alpha=0,\\
1\quad&\cos\alpha=1/\sqrt{d},\\
3/2\quad&\cos\alpha=\sqrt{3/(d+2)}.
\end{cases}
\end{equation}

\section{Scaling Properties}

The special values listed in \eqref{beta-special} suggest that the
survival exponent has the scaling form
\begin{equation}
\label{beta-y-scaling}
\beta_d(\alpha)\to \beta(y) \quad {\rm with}\quad y=(\cos\alpha)\sqrt{d},
\end{equation}
in the limit $d\to\infty$. From equation \eqref{beta-special}, we
deduce \hbox{$\beta(y=0)=1/2$}, \hbox{$\beta(y=1)=1$}, and
\hbox{$\beta(y=\sqrt{3})=3/2$}.  To show the scaling behavior
\eqref{beta-y-scaling} in general, we introduce the variable
$z=\mu\,\sqrt{d}$. Performing this scaling transformation on the
Laplace equation \eqref{psi-mu-eq} shows that the angular function
depends on a single scaling variable, $\psi_d(\mu)\to \psi(z)$, in the
$d\to\infty$ limit. The scaling function $\psi(z)$ satisfies
\begin{equation}
\label{psi-z-eq}
\psi_{zz}-z\,\psi_z+2\beta \psi=0.
\end{equation}
The boundary condition is $\psi(z=y)=0$, and additionally, $\psi(z)>0$
for all $y<z<\infty$.

\begin{figure}[t]
\includegraphics[width=0.45\textwidth]{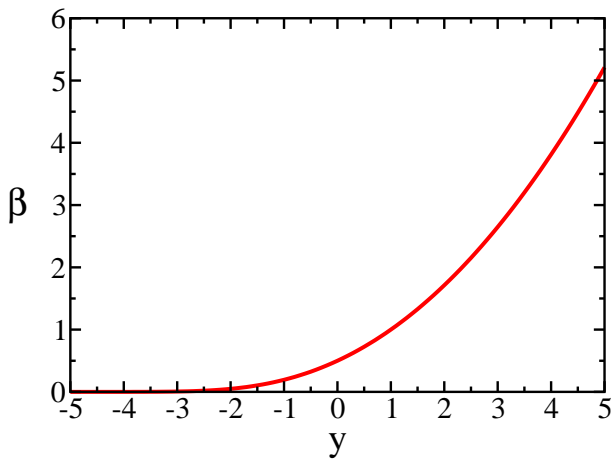}
\caption{The exponent $\beta$, specified by equation \eqref{beta-y},
versus the scaling variable $y=(\cos\alpha)\sqrt{d}$.}
\label{fig-scaling}
\end{figure}

Using the transformation $\psi(z)=\exp(z^2/4)u(z)$, we recast
\eqref{psi-z-eq} into the parabolic cylinder equation \cite{bo}
\begin{equation*}
u_{zz}+\left(2\beta+\frac{1}{2}-\frac{z^2}{4}\right)u=0.
\end{equation*}
The two independent solutions of this equation are $D_{2\beta}(z)$ and
$D_{2\beta}(-z)$ where $D_\nu(z)$ is the parabolic cylinder function
of index $\nu$. The second solution is not physical: the asymptotic
behavior \hbox{$D_\nu(-z)\sim \exp(z^2/4)$} as $z\to\infty$ \cite{bo}
implies a divergent survival probability in the limit $\theta\to
0$. Therefore,
\begin{equation}
\label{psi-z}
\psi(z)=e^{z^2/4}D_{2\beta}(z).
\end{equation}
As always, we set the overall proportionality constant to one.  The
boundary condition $\psi(z=y)=0$ relates the scaling function
$\beta(y)$ and the scaling variable $y$, defined in \eqref{beta-y-scaling},
 \begin{equation}
\label{beta-y}
D_{2\beta}(y)=0.
\end{equation}
The proper solution is the largest root of the parabolic cylinder
function. For half-integer values of the exponent, the parabolic
cylinder function is related to a Hermite polynomial,
$D_{2\beta}(z)\propto H_n(z/\sqrt{2})\exp(-z^2/4)$, and hence,
\eqref{beta-y} is equivalent to $H_n(y/\sqrt{2})=0$, where the largest
root is the appropriate one.  Therefore, the half-integer values of
the scaling function $\beta(y)$ occur at zeroes of Hermite
polynomials.  In addition to the aforementioned values, we also quote
$\beta\big(\sqrt{3+\sqrt{6}}\,\big)=2$ and
$\beta\big(\sqrt{5+\sqrt{10}}\,\big)=5/2$.

The scaling behavior \eqref{beta-y-scaling} is valid in the limits
\hbox{$\alpha\to\pi/2$} and $d\to\infty$ with the scaling variable
$y=(\cos\alpha)\sqrt{d}$ or alternatively $y=(\pi/2-\alpha)\sqrt{d}$
kept finite. Hence, the scaling function in \eqref{beta-y} quantifies
the shape of $\beta$ in a ``scaling window'' \cite{bbckw,bk1,abk}
centered on $\alpha=\pi/2$. The size of this window shrinks as
$d^{-1/2}$ at large dimensions.  As shown in Figure \ref{fig-scaling},
$\beta$ vanishes as \hbox{$y\to-\infty$}, and it diverges as
\hbox{$y\to\infty$}. Surprisingly, if we fix the opening angle
$\alpha$ and then take the $d\to\infty$ limit, there are three
distinct possibilities, as stated in \eqref{three}: the exponent
vanishes if $\alpha>\pi/2$, it always equals $1/2$ when
$\alpha=\pi/2$, and it diverges if $\alpha<\pi/2$.

\begin{figure}[t]
\includegraphics[width=0.45\textwidth]{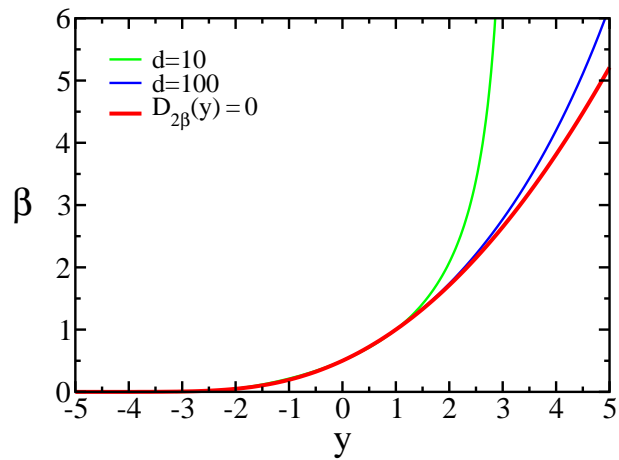}
\caption{The convergence to the scaling behavior. The exponent
$\beta\equiv \beta_d(\alpha)$, specified by \eqref{betad}, is plotted
versus the scaling variable $y=(\cos\alpha)\sqrt{d}$ for $d=10$ and
$d=100$. Also shown is the scaling function $\beta(y)$.}
\label{fig-convergence}
\end{figure}

We also note that the approach to the scaling behavior is not
uniform.  The scaling exponent $\beta$ converges rapidly to the
scaling function for negative $y$, but the convergence is quite slow
for positive $y$ (Figure \ref{fig-convergence}).

To find the asymptotic behavior when $y\to\infty$, we use 
the fact that the largest root $\xi\equiv \xi(\nu)$ of the parabolic
cylinder function of index $\nu$, $D_\nu(\xi)=0$, is located at
\cite{as}
\begin{equation*}
\xi\simeq (4\nu)^{1/2}-|a_1|\,\nu^{-1/6},
\end{equation*}
when $\nu\to\infty$.  Here, $a_1\cong -2.338107$ is the first root of
the Airy function. Substituting $\nu=2\beta$ into this expression, we
find $y\simeq (8\beta)^{1/2}-|a_1|(2\beta)^{-1/6}$ as
$\beta\to\infty$.  Therefore, the leading behavior is $\beta\simeq
y^2/8$. Furthermore, the first correction to this asymptotic form is
\begin{equation}
\label{beta-y-large}
\beta\simeq \frac{y^2}{8}+|a_1|2^{-5/3}\,y^{2/3}.
\end{equation}
Thus, the exponent $\beta$ diverges algebraically when
\hbox{$y\to\infty$}.

We use perturbation analysis to find how $\beta$ vanishes in the
complementary limit $y\to\infty$.  Equation \eqref{psi-z-eq} shows
that the angular function becomes constant when $\beta\to
0$. Thus, $\psi(z)\simeq 1+\beta\, g(z)$.  Substituting this form into
\eqref{psi-z-eq}, the correction function $g\equiv g(z)$ obeys
\begin{equation}
\label{g-z-eq}
g_{zz}-z g_z+2=0. 
\end{equation}
The boundary condition 
\begin{equation}
\label{g-z-bc}
g_z(0)=\sqrt{2\pi}
\end{equation}
follows from the small-$\beta$, small-$z$, behavior of the scaling solution
\eqref{psi-z}
\begin{equation*}
\psi(z)\simeq
1+\frac{D_{2\beta}'(0)}{D_{2\beta}(0)}\,z =  
1-\frac{\sqrt{2}\,\Gamma(\tfrac{1}{2}-\beta)}{\Gamma(-\beta)}\,z
\simeq 1+\sqrt{2\pi}\,\beta\, z.
\end{equation*}
Here, we used the values 
\hbox{$D_{2\beta}(0)=\sqrt{\pi}\,2^\beta/\Gamma(1/2-\beta)$} and
\hbox{$D_{2\beta}'(0)=-\sqrt{\pi}\,2^{\beta+1/2}/\Gamma(-\beta)$}
\cite{bo}, as well as the identities $\Gamma(1/2)=\sqrt{\pi}$ and
$(-\beta)\Gamma(-\beta)=\Gamma(1-\beta)$.  

Using the integrating
factor \hbox{$I(z)=\exp(-z^2/4)$}, we simplify
Eq.~\eqref{g-z-eq} to $d(g_z\,I)/dz=-2I$. Integration of this equation
subject to the boundary condition \eqref{g-z-bc} gives
\begin{equation}
g(z)=2 \int_0^z dt\, e^{t^2/2}\int_t^\infty ds\, e^{-s^2/2}.
\end{equation}
The second integral approaches a constant, and consequently,
\hbox{$g(z)\simeq -\sqrt{8\pi/z^2}\exp(z^2/2)$}, when \hbox{$z\to
-\infty$}. We now impose the boundary condition
\hbox{$\psi(y)=1+\beta\, g(y)=0$}, and find that the exponent is
exponentially small,
\begin{equation}
\label{beta-y-small}
\beta(y)\simeq \sqrt{\frac{y^2}{8\pi}}\,e^{-y^2/2},
\end{equation}
in the limit $y\to-\infty$.

In summary, the scaling function $\beta(y)$ has the following extremal
behaviors
\begin{equation}
\label{beta-y-limits}
\beta(y)\simeq
\begin{cases}
\sqrt{y^2/8\pi}\,\exp\!\left(-y^2/2\right)&y\to-\infty,\\
y^2/8 & y\to\infty.
\end{cases}
\end{equation}
At large dimensions, these limiting behaviors apply only inside the
scaling window, that is, when $|\pi/2-\alpha|$ is of the order
$d^{-1/2}$. In the next two sections, we perform asymptotic analysis
for the limiting cases of extremely thin cones ($\theta\to 0$) and 
extremely wide cones ($\theta\to\pi)$.

\section{Thin Cones}

The explicit expressions \eqref{beta2} and \eqref{beta4} show that the
exponent is inversely proportional to the opening angle, $\beta\sim
\alpha^{-1}$, when the cone is extremely thin. We anticipate that this
divergence is generic.

When the cone is very thin, we have \hbox{$\sin\theta\simeq \theta$}
and equation \eqref{psi-theta-eq} simplifies,
\begin{equation}
\label{psi-theta-eq-thin}
\frac{d^2 \psi}{d \theta^2} + 
\frac{d-2}{\theta}\,\frac{d \psi}{d \theta}+(2\beta)^2 \psi=0.
\end{equation}
This equation holds as long as $d\alpha^2\ll 1$. In writing this
equation we tacitly assumed a divergent $\beta$.  We now introduce the
scaling variable $x=2\beta\theta$ and transform equation
\eqref{psi-theta-eq-thin} as follows,
\begin{equation*}
\psi_{xx} + \frac{d-2}{x}\,\psi_x+\psi=0.
\end{equation*}
Next, we seek a solution in the form $\psi(x)=x^{-\delta}u(x)$, where
as in \eqref{psid}, $\delta=\frac{d-3}{2}$.  Performing this second
transformation gives the Bessel equation \cite{jdw}
\begin{equation*}
x^2\,u_{xx}+x\,u_x+\left(x^2-\delta^2\right)u=0.
\end{equation*}
This equation has two independent solutions: the Bessel functions
$J_\delta(x)$ and $Y_\delta(x)$.  The boundary condition $\psi'(0)=0$
excludes the latter and hence,
\begin{equation}
\label{psi-thin}
\psi(x)= x^{-\delta}J_\delta(x).
\end{equation}
The other boundary condition, $\psi(\alpha)=0$, gives an implicit
relation $J_\delta(2\beta\alpha)=0$ between the exponent $\beta$ and
the opening angle $\alpha$. We see that the exponent is inversely
proportional to the opening angle in all dimensions:
\begin{equation}
\label{beta-thin}
\beta_d(\alpha)\simeq B_d\,\alpha^{-1}\quad {\rm with}\quad
B_d=\tfrac{1}{2}\zeta(\delta),
\end{equation}
as $\alpha\to 0$.  Here, $\zeta(\delta)$ is the first positive zero of
the Bessel function, $J_\delta(\zeta)=0$.  Table I lists the coefficients $B_d$
for $d\leq 10$.

\begin{table}[t]
\begin{tabular}{|c|l|}
\hline
$d$&$B_d$\\
\hline
$2$ & $0.785398$ \\
$3$ & $1.202412$ \\
$4$ & $1.570796$ \\
$5$ & $1.915852$ \\
$6$ & $2.246704$ \\
$7$ & $2.567811$ \\
$8$ & $2.881729$ \\
$9$ & $3.190080$ \\
$10$ & $3.493966$ \\
\hline
\end{tabular}
\caption{The coefficient $B_d$ in \eqref{beta-thin} for $d\leq
10$. Equations \eqref{psi2} and \eqref{psi4} yield $B_2=\pi/4$ and
$B_4=\pi/2$, respectively.}
\end{table}

At large dimensions, we use the asymptotic behavior \cite{as,NIST}
\begin{equation*}
\zeta(\delta)\simeq \delta+|a_1|(\delta/2)^{1/3}
\end{equation*}
when $\delta\to\infty$. Again, $a_1\cong -2.338107$ is the first root
of the Airy function. Thus, to leading order, the prefactor $B_d$ is
linear in the dimension, $B_d\simeq d/4$. The correction to the
leading asymptotic behavior is given by
\begin{equation}
\label{Bd-larged}
B_d\simeq \frac{d}{4}+|a_1|2^{-5/3}d^{1/3},
\end{equation}
for $d\to\infty$. The divergence \eqref{beta-thin} shows that a
diffusing particle quickly reaches the cone surface when the cone is
thin. Moreover, equation \eqref{Bd-larged} implies $\beta\simeq
d/4\alpha$, and hence, the first-passage process becomes even faster 
as the dimension increases.

\section{Wide Cones}

For all $d\geq 3$, the solutions to equation \eqref{betad} show that
the exponent vanishes when the interior of the cone occupies all of
space (Figure \ref{fig-beta}).  In this case, equation
\eqref{psi-theta-eq} becomes
\begin{equation}
\label{psi-theta-eq-wide}
\frac{1}{(\sin\theta)^{d-2}}\frac{d}{d \theta}\left[(\sin\theta)^{d-2}
\frac{d\psi}{d  \theta}\right]+2\beta(d-2)\psi=0.
\end{equation}
We obtain $\beta$ by repeating the perturbation analysis leading to
\eqref{beta-y-small}. For all $d\geq 3$, we write $\psi(\theta)\simeq
1+\beta\,g(\theta)$.  The boundary condition $\psi'(0)=0$ implies
$g'(0)=0$.  From \eqref{psi-theta-eq-wide}, the correction $g\equiv
g(\theta)$ obeys
\begin{equation}
\label{g-eq-theta-1}
\frac{d}{d \theta}\left[(\sin\theta)^{d-2}
\frac{dg}{d\theta}\right]=-2(d-2)(\sin\theta)^{d-2}.
\end{equation}
Integrating this equation twice and using the boundary condition
$g'(0)=0$, we find the correction up to a constant,
\begin{equation}
\label{double-int}
g(\theta)=g_0-2(d\!-\!2)\!\int_0^\theta\!\!\!
\frac{d\phi}{(\sin\phi)^{d-2}}\int_0^\phi \!\! d\varphi\, (\sin\varphi)^{d-2}.
\end{equation}
In the limit $\theta\to\pi$, the second integral approaches the constant
\begin{equation*}
S_d=\int_0^\pi d\theta \, (\sin\theta)^{d-2}=
\frac{\Gamma\left(\tfrac{d-1}{2}\right)
\Gamma\left(\tfrac{1}{2}\right)}{\Gamma\left(\tfrac{d}{2}\right)}.
\end{equation*}
Moreover, the term $(\sin\phi)^{-(d-2)}\simeq (\pi-\phi)^{-(d-2)}$
dominates the first integral and therefore, the correction function 
diverges,
\begin{equation*}
g(\theta)\simeq -\frac{2(d-2)S_d}{d-3}(\pi-\theta)^{-(d-3)},
\end{equation*}
as $\theta\to\pi$. We now impose the boundary condition
$\psi(\alpha)=1+\beta g(\alpha)=0$ and find that the exponent $\beta$
vanishes exponentially as $\alpha\to\pi$,
\begin{equation}
\label{beta-wide}
\beta_d(\alpha)\simeq A_d\,(\pi-\alpha)^{d-3}. 
\end{equation}
The coefficient $A_d=\frac{d-3}{2(d-2)S_d}$ is given by
\begin{equation}
\label{Ad}
A_d=\frac{\Gamma\left(\tfrac{d-2}{2}\right)}
{2\,\Gamma\left(\tfrac{1}{2}\right)\Gamma\left(\tfrac{d-3}{2}\right)}.
\end{equation}
In particular $A_4=1/2\pi$, in agreement with equation
\eqref{beta4}. Table II lists the coefficients $A_d$ for $d\leq 10$.
The asymptotic property $\Gamma(x+a)/\Gamma(x)\to x^a$ as $x\to\infty$
shows that the coefficient grows algebraically with dimension,
$A_d\simeq \sqrt{d/8\pi}$.

\begin{table}[t]
\begin{tabular}{|c|c|c|c|c|c|c|c|c|}
\hline
$d$&$3$&$4$&$5$&$6$&$7$&$8$&$9$&$10$\\
\hline 
$A_d$&
$\ 0\ $& $\ \frac{1}{2\pi}$&
$\ \frac{1}{4}\ $&$\ \frac{1}{\pi}\ $&
$\ \frac{3}{8}\ $&$\ \frac{4}{3\pi}\ $&
$\ \frac{15}{32}\ $&$\ \frac{8}{5\pi}\ $\\
[2pt]
\hline
\end{tabular}
\caption{The coefficient $A_d$ given by equation \eqref{Ad}
for \hbox{$d\leq 10$}.}
\end{table}

In the marginal case $d=3$, the coefficient in \eqref{beta-wide}
vanishes, $A_3=0$.  In this case, the exponent vanishes gently,
\begin{equation}
\beta_3\simeq \frac{1}{4\ln \frac{2}{\pi-\alpha}}
\end{equation}
as follows from \eqref{beta3} \cite{jdw}. This behavior is consistent
with the near-$\pi$ behavior of \eqref{double-int}, $g(\theta)\simeq
-4\ln\frac{1}{\pi-\theta}$.

As the opening angle approaches its maximal value, the cone surface
turns into an infinitely long, yet infinitesimally thin needle.  In
two dimensions, a diffusing particle is bound to reach such a
needle. Yet, in dimensions three and higher, the particle may or may
not reach the needle. Moreover, the exponentially small exponent
\eqref{beta-wide} shows that the first-passage process becomes
extremely slow at high dimensions.

\section{First-Passage Time}

We now briefly discuss the first-passage time. Let $T(r,\theta)$ be
the average duration of the first-passage process, namely, the average
time it takes a particle released at $(r,\theta)$ to reach the cone
surface for the first time (Figure \ref{fig-cone}). This quantity
obeys the Poisson equation \cite{sr}
\begin{equation}
\label{T-eq}
D\nabla^2 T(r,\theta) = -1,
\end{equation}
and the boundary condition $T(r,\alpha)=T(0,\theta)=0$. As in
\eqref{Phi-sep}, we use dimensional analysis and write \hbox{$T
(r,\theta)= (r^2/D)\,U(\theta)$} where $U(\theta)$ is a dimensionless
function of the angle. From Eq.~\eqref{T-eq}, we get
\begin{equation}
\label{U-eq}
\frac{1}{(\sin\theta)^{d-2}}
\frac{d}{d \theta}\left[(\sin\theta)^{d-2}\frac{d U}{d \theta}\right]
+ 2d U = -1, 
\end{equation}
while the boundary condition becomes $U(\alpha)=0$. The linear
equation \eqref{U-eq} has the particular solution $\tilde U=-1/2d$,
and thus, we seek a solution in the form
\hbox{$U(\theta)=u(\theta)-1/2d$}.  The function $u\equiv u(\theta)$
obeys the homogeneous linear differential equation
\begin{equation*}
\frac{1}{(\sin\theta)^{d-2}} \frac{d}{d
\theta}\left[(\sin\theta)^{d-2}\frac{d u}{d \theta}\right] + 2du = 0,
\end{equation*}
subject to the boundary condition $u(\alpha)=1/2d$.  This equation is
a special case of \eqref{psi-theta-eq}, and using \eqref{psi-special},
we immediately find \hbox{$u(\theta)=C(d\cos^2\theta - 1)$} where $C$
is set by the boundary condition. Finally, we find that the
first-passage time has the compact form
\begin{equation}
\label{T}
T(r,\theta)=\frac{r^2}{2D}\,
\frac{\cos^2\theta-\cos^2\alpha}{d\cos^2\alpha-1}.
\end{equation}
This first-passage time is finite if and only if $\beta>1$.

\section{Discussion}

In summary, we studied first-passage kinetics for a particle diffusing
in a cone. In all dimensions, the probability that the particle does
not reach the cone boundary decays algebraically with time.  We found
the exponent underlying this power-law behavior as a root of a
transcendental equation involving associated Legendre functions. We
also obtained scaling and extremal properties of the exponent.

Our results generalize the known properties of first passage in two-
and three-dimensional cones \cite{sr} to arbitrary dimensions.
Moreover, the statistical physics perspective, where scaling plays a
central role, extends rigorous studies of diffusion in cones in
probability theory \cite{rdd,dz,bs,bd} and potential theory
\cite{ntv}.  Scaling implies that the behavior in a narrow window
becomes universal: cones with different combinations of $\alpha$ and
$d$, have the same exponent $\beta$, as long as the scaling variable
$y=(\cos\alpha)\sqrt{d}$ is the same.  The exponent is exponentially
small when $y\to-\infty$, and it grows algebraically when
$y\to\infty$.

As the dimension increases, the first-passage process speeds up if the
opening angle is acute but slows down if the opening angle is
obtuse. This behavior is reflected by the asymptotic behavior for
fixed $\theta$ in the large-$d$ limit.  We merely quote the results of
the corresponding asymptotic analysis,
\begin{equation}
\label{beta-larged}
\beta_d(\alpha)\simeq 
\begin{cases}
\frac{d}{4}\left(\frac{1}{\sin\alpha}-1\right) & \alpha<\pi/2,\\
C(\sin\alpha)^d& \alpha>\pi/2.
\end{cases}
\end{equation}
The exponent is proportional to the dimension at acute angles, and it
decays exponentially with dimension at obtuse angles.  The asymptotic
behavior \eqref{beta-larged} is consistent with the limiting behavior
of the scaling function \eqref{beta-y-limits} as well as the
asymptotic results \eqref{beta-thin} and \eqref{beta-wide}. 

Kinetics of first-passage in a cone provide useful information about
first-passage problems involving multiple random walks. For example,
let us consider the random-walk problem mentioned in the introduction.
The probability that the trajectories of $N$ ordinary random walks do
not meet up to time $t$ is equivalent to the probability that a
compound random walk in $N$ dimensions remains inside the region
\hbox{$x_1<x_2<\cdots<x_N$}. Since there are $N!$ permutations of the
positions, the total solid angle inside this ``allowed region''
accounts for a fraction $1/N!$ of space. We replace the allowed region
with an $N$-dimensional cone that has the same solid angle,
$\alpha^{N-1}\sim 1/N!$. Using the Stirling formula, \hbox{$N!\sim
\sqrt{2\pi N}(N/e)^N$}, the opening angle is $\alpha\simeq e/N$.  The
survival probability decays algebraically, $S(t)\sim t^{-\beta_N}$,
and using \eqref{beta-thin}--\eqref{Bd-larged}, the exponent is
$\beta_N\simeq N^2/4e$.  This cone approximation correctly gives the
asymptotic $N$-dependence as the exact exponent is $N(N-1)/4$
\cite{mef,hf,djg}.

First passage in the exterior of this narrow cone gives an
approximation for the probability that the order of $N$ random walks
does not turn into the mirror image of the initial state.  If the
initial ordering of the random walks is \hbox{$\{1,2,\cdots,N-1,N\}$},
then $S(t)$ is the probability that the particles do not reach the
configuration \hbox{$\{N,N-1,\ldots,2,1\}$} up to time $t$. Since the
angle of the cone is now $\pi-\alpha\simeq e/N$, the asymptotic
behavior \eqref{beta-wide} gives $S(t)\sim t^{-\beta_N}$ with the tiny
exponent
\begin{equation*}
\beta_N\simeq \sqrt{N/8\pi}\,(e/N)^{N-3}.
\end{equation*}
In a follow-up study, we use cones to understand other random walk
problems \cite{bk}.
 
With a few exceptions such as paraboloids \cite{bds,ls}, the study of
first passage in elementary geometries is still in its infancy.  We
considered first passage in basic circular cones. A natural extension
is to generalized cones \cite{dlb}, defined as domains with the
property that all rays emanating from the apex do not intersect the
cone boundary.  Understanding first-passage properties in such
generalized cones ultimately requires a solution to the
Laplace-Dirichlet boundary value problem.

\medskip
We thank Sidney Redner for useful discussions. This
research has been supported by DOE grant DE-AC52-06NA25396 and NSF
grant CCF-0829541.

\end{document}